\documentclass[12pt]{article}
\usepackage{epsfig}
\usepackage{cite}
\usepackage{./Macros/mcite}
\usepackage{url}

\chardef\til=126
\newcommand{\mev}{{\,\mathrm{MeV}}}
\newcommand{\gev}{{\,\mathrm{GeV}}}
\newcommand{\tp}{\Theta^+ (1530)}
\newcommand{\ks}{K^{0}_{S}}
\newcommand{\ksp}{K^{0}_{S}\> p}
\newcommand{\kspb}{K^{0}_{S}\> \bar{p}}
\newcommand{\ee}{e^+e^-}

\begin{document}

\clearpage
\pagestyle{empty}
\setcounter{footnote}{0}\setcounter{page}{0}%
\thispagestyle{empty}\pagestyle{plain}\pagenumbering{arabic}%

\hfill ANL-HEP-PR-05-06
 
\hfill February  2005

% \hfill version 2.0  

\vspace{2.0cm}
 
\begin{center}
 
%%%%%%%%%%%%%%%%%%%%%%%%%%%%%%%%%%%%%%%%%%%%%%%%%%%%%%%%%%%%%%%
{\Large\bf  
Pentaquarks in high-energy colliding experiments:  
perspectives from HERA\\[-1cm] }
%%%%%%%%%%%%%%%%%%%%%%%%%%%%%%%%%%%%%%%%%%%%%%%%%%%%%%%%%%%%%%%

\vspace{2.5cm}

{\large S.~Chekanov}

\vspace{0.5cm}

HEP division, Argonne National Laboratory, \\
9700 S.Cass Avenue,
Argonne, IL 60439
USA \\ 
E-mail: chekanov@mail.desy.de

\vspace{0.5cm}
\begin{abstract}
Several issues related to pentaquark searches relevant
for current and future high-energy colliding experiments are discussed. 
We make an attempt to explain why pentaquark candidates 
are not seen by some experiments, and what makes the HERA 
experiments so special in such searches.   
\end{abstract}

\end{center}

\newpage
\setcounter{page}{1}

%%%%%%%%%%%%%%%%%%%%%%%%%%%%%%%%%%%%%%%%%%%%%%%%%%%%%%%%%%%%%%%%%%
\section{Introduction}
%%%%%%%%%%%%%%%%%%%%%%%%%%%%%%%%%%%%%%%%%%%%%%%%%%%%%%%%%%%%%%%%%%
Recently, many experimental groups made significant efforts to find  
new states which can be explained as consisting of five quarks.
A number of experiments~\cite{Nakano:2003qx,*Barth:2003ja,*Stepanyan:2003qr,*prl92:032001,*Barmin:2003vv,*Aleev:2004sa,*hep-ex-0403011,Airapetian:2003ri} 
including ZEUS~\cite{Chekanov:2004kn}
have reported narrow signals in the vicinity of $1530\mev$
in the $nK^{+}$ and $\ksp$ invariant-mass 
spectra.
The signals were  consistent with the exotic 
pentaquark baryon state $\tp$
with quark content $uudd\bar{s}$ \cite{zp:a359:305}.
Several other experiments searched for this state with negative
results~\cite{Bai:2004gk,*Abt:2004tz,*Antipov:2004jz,pl:599:1,cdfpenta,Aubert:2004bm,Abe:2004ws}.
ZEUS also reported negative searches for other pentaquark 
candidates~\cite{Chekanov:2004qm,Chekanov:2004ku,Chekanov:2004hd,*Karshon:2004kt}.

At present, ZEUS is the only high-energy colliding 
experiment which observes the 
$\tp$ state in the $\ksp$ decay channel. 
The H1 experiment has evidence for the lightest charmed pentaquark
state~\cite{Aktas:2004qf}.   
Several other colliding experiments, 
BaBar~\cite{Aubert:2004bm}, ALEPH~\cite{pl:599:1}, Belle~\cite{Abe:2004ws} and  CDF~\cite{cdfpenta} 
do not observe any of the pentaquarks.
Although in some cases the results are still preliminary,  
it is evident that the statistics of such experiments 
for several known states exceed the statistics
of those experiments which observe the pentaquarks. 
While non-observation of pentaquarks
in $\ee$ collisions does not necessarily contradict to the
observation of a signal predominantly
produced in the forward region
with a baryon in the beam or in the target,
it is more difficult to explain why no pentaquarks are seen in
$p\bar{p}$  collisions at Tevatron and for some other fixed-target experiments.

The aim of this article is not to give a detailed review of the present
status of pentaquark searches, which can be found 
elsewhere~\cite{Dzierba:2004db,Hicks:2004ge,*Kabana:2005kg}, but
to discuss the most intriguing problem: 
why the ZEUS experiment observes the  $\tp$ state in deep inelastic scattering (DIS) 
at medium $Q^2$. The answer on this question may also explain the  
negative searches at Tevatron, and perhaps, the negative results for some fixed-target experiments.
We also will discuss several experimental issues which could be 
important for current and future
pentaquark searches in high-energy experiments.  

The existing experimental results are still rather controversial, 
therefore, it is difficult to discuss them  without
making any assumptions. 
We assume that the $\tp$ state in the  $\ksp$ decay mode 
is indeed a pentaquark, but not a new 
$\Sigma$ state with a strongly suppressed $\Lambda\pi$ decay mode.  
Secondly, at present, there is rather weak 
evidence for antipentaquarks: the ZEUS result~\cite{Chekanov:2004kn}  indicates
that the peak at $1522\mev$ is mainly driven by the $\ksp$ 
decay mode rather than due to   
the $\kspb$ channel 
(although a structure at $1480\mev$, 
which is consistent with the PDG $\Sigma(1480)$ bump,  is more 
pronounced for the $\kspb$ invariant mass).
Since antipentaquarks can only be produced by soft 
fragmentation, the second observation
avoids the contradiction with negative searches in 
$\ee$ annihilation experiments.

%%%%%%%%%%%%%%%%%%%%%%%%%%%%%%%%%%%%%%%%%%%%%%%%%%%%%%%%%%%%%%%%%%
\section{Possible production mechanism}
\label{prod}
%%%%%%%%%%%%%%%%%%%%%%%%%%%%%%%%%%%%%%%%%%%%%%%%%%%%%%%%%%%%%%%%%%

We will start with the discussion of possible differences in the production
mechanism of baryonic states in $ep$ collisions at 
HERA and other experiments.

Probably, the most simple explanation of why 
pentaquarks have not been observed in $\ee$
experiments lays in the fact that 
these experiments do not have the net baryon number in the colliding
beams. If one thinks in terms of the pentaquark 
explanation of the $\tp$ signal,
the production of this  state in $\ee$ should be 
significantly suppressed 
since it requires
five quarks to be glued together by the soft 
fragmentation mechanism - something which is difficult to explain using the 
conventional hadronisation mechanisms.
This could  explain the negative results in $\ee$ annihilation experiments.

The situation is different when a baryonic  state is 
present in either beam and/or
in the target. In this case, 
the fragmentation mechanism is not as simple as in $\ee$, and 
it is possible that the hadronic final state would have  
enhanced baryon production rate in comparison with the $\ee$ annihilation.  
If the presence of the net baryon number is indeed a 
necessary condition for
formation of the $\tp$ pentaquark, then one should 
explain the negative results of the CDF experiment \cite{cdfpenta}.

It should be emphasised that the $\tp$ state was 
observed in DIS
at the exchanged photon virtuality $Q^2>20\gev^2$ \cite{Chekanov:2004kn}, 
while no signal was reported in photoproduction ($Q^2\simeq 0$). 
The photoproduction events are  dominated by the boson-gluon 
fusion (BGF) mechanism, $\gamma^{*} g\to q\bar{q}$, which makes
the $ep$ collisions to be very similar to $p\bar{p}$. 
Thus the negative result from Tevatron 
is not in contradiction with  non-observation of the $\tp$ state
in photoproduction at HERA. 

Note that the photoproduction 
data selected by ZEUS are trigger-biased mainly by the detection
of high $E_T$ jets. 
Therefore, such data are enriched  with the
BGF-type of events.
In contrast, DIS data are free from such bias as they were
collected by triggering the scattered electron, 
with a minimum bias on the hadronic final 
state.   

What makes the HERA data so special with respect to exotic baryon searches?
In $ep$ collisions, a significant fraction of events contain a leading
baryon carrying a large fraction of the
incoming proton momentum~\cite{Adloff:1998yg,Adloff:2001be,Chekanov:2002pf,Chekanov:2002yh}.
There are several models describing the leading baryon production.
One type of models is based on the QCD building blocks, quarks and gluons,
and explains such baryons as the result of the hadronisation
of the spectators from the incoming proton. In the second type of models, 
the leading baryons can be explained by the exchange of virtual particles.
Finally, such baryons could originate from decays of low-mass proton
resonances in diffractive interactions.

Recently, both H1 \cite{Adloff:2001be} and 
ZEUS~\cite{Chekanov:2002pf} reported a 
rise in the leading-baryon yield with increase of $Q^2$ 
from the photoproduction regime to DIS. This could be explained by absorptive
effects. 
If one assumes that a simitar effect could also  contribute to the exotic-baryon
production, this may explain    
why ZEUS does not see the $\tp$ signal for 
photoproduction and low $Q^2$ DIS.
This could be a good guess, if one proves that the most
central fragmentation region, corresponding to the
pseudorapidity range  $\mid \eta \mid<1.5$ used for   
ZEUS pentaquark searches,  can also be populated
with baryons after fragmentation 
of the nucleon remnant (but, of course, with a significantly smaller probability
than that for the most forward $\eta$ region 
accessed by the leading-baryon calorimeters
installed by ZEUS and H1).

To verify this assumption, Monte Carlo (MC) studies were
performed using the following models:               
PYTHIA6.2~\cite{Sjostrand:2001yu}, 
HERWIG 6.5~\cite{herwig}, 
ARIADNE 4.12~\cite{Lonnblad:1992tz} and 
LEPTO 6.5~\cite{Ingelman:1996mq}.
The physics calculations were performed
with the module ``baryons\_DIS.rmc'' of the RunMC package \cite{runmc}. 
The MC models do not contain the simulation of the leading-baryon effects
due to the exchange of virtual particles, but rather they
attempt to describe the effect assuming that the nucleon remnant
is a diquark composed of the valence quarks as spectators.
While PYTHIA, ARIADNE and LEPTO use the LUND model to describe 
the hadronisation by stretching a colour triplet string between
the struck quark and the diquark\footnote{For BGF, this picture is modified
as tree quarks remain in colour octet after the gluon emission.}, 
the HERWIG model uses the cluster
model to hadronise such diquarks. In case of LEPTO,
the soft-colour interaction model is used to  
simulate the rapidity-gaps events.

At this point, we are not interested in how realistic such models
are in the description of the production rate of the leading baryons 
which carry a large fraction of the incoming proton
energy~\cite{Adloff:2001be,Chekanov:2002pf}.   
The main question is whether they can predict events which have 
an enhanced production rate of the conventional baryons,    
compared to that of antibaryons 
for the central fragmentation region where the $\tp$ state
was seen by ZEUS.

\begin{figure}
\begin{center}
\mbox{\epsfig{file=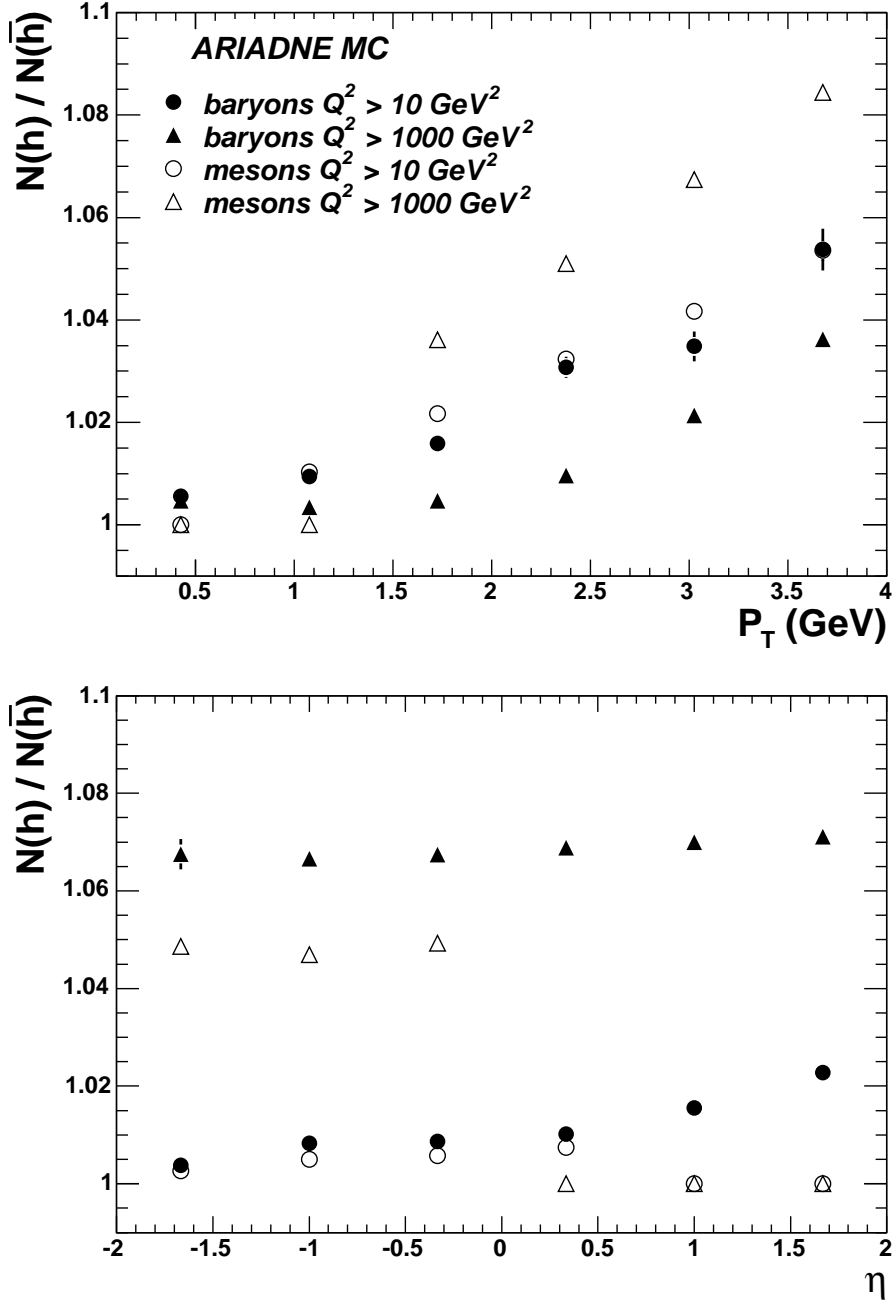,height=18cm}}
\caption
{
The ratio of the total number of final-state hadrons to antihadrons, $N(h)/N(\bar{h})$, 
as functions of $p_T$ and $\eta$ in DIS. The ARIADNE Monte Carlo model is used. 
The ratios are shown for mesons and baryons separately for two regions in
$Q^2$. 
}
\label{r_dis}
\end{center}
\end{figure}

Figure~\ref{r_dis} shows the hadron-antihadron
asymmetry separately for mesons and baryons
as functions of $p_T$ and $\eta$ in the laboratory frame. 
The pseudorapidity was defined as $\eta=-\ln\left(\tan\frac{\theta}{2}\right)$,
where the polar angle, $\theta$, is measured with respect to the proton beam
direction (i.e. positive $\eta$ corresponds to the proton beam direction).
The Monte Carlo simulations were performed with ARIADNE
for $Q^2>10\gev^2$ and $Q^2>1000\gev^2$. 
The simulations were done for the proton 
($E_p=920\gev$) and the electron ($E_e=27.5\gev$) colliding beams.
The asymmetry was defined as  $r(h)=N(h)/N(\bar{h})$, where $N(h)$ 
($N(\bar{h})$) is the number of counted hadrons (antihadrons) and $h=m,b$,
where $m$ and $b$  denote mesons and baryons, respectively.  
The minimum transverse momentum for hadrons was set to $150\mev$ as for the
experimental searches~\cite{Chekanov:2004kn}. 

It is interesting to observe that the ratios shown in Fig.~\ref{r_dis} are
above unity, i.e. they indicate the asymmetry.   
This is not a surprise since 
the hadron-final state is affected by the presence of the valence 
quarks from the incoming proton with the total electric charge $+1$. 
A similar effect is also expected for all types of hadrons (mesons and baryons) 
in the current jet~\cite{hep-ph/0003255},
and it was found to be stronger with increase of $Q^2$.    
The most interesting feature of Fig.~\ref{r_dis} 
is that the asymmetry is larger for baryons than for mesons,
and this tendency increases at low $p_T$ and positive $\eta$. 
This could be an indication of the presence of the proton debris after
the fragmentation process. 

Thus, occasionally, baryons    
can be  kicked out of a proton by the fragmentation process
to the central pseudorapidity region. The probability
of such effect is small, but not negligible, especially at high $Q^2$.
As was discussed above, there are other mechanisms 
responsible for the leading-baryon production, and they may further contribute
to the baryon-anibaryon asymmetry  in 
the pseudorapidity region shown in Fig.~\ref{r_dis}.  

\begin{figure}
\begin{center}
\mbox{\epsfig{file=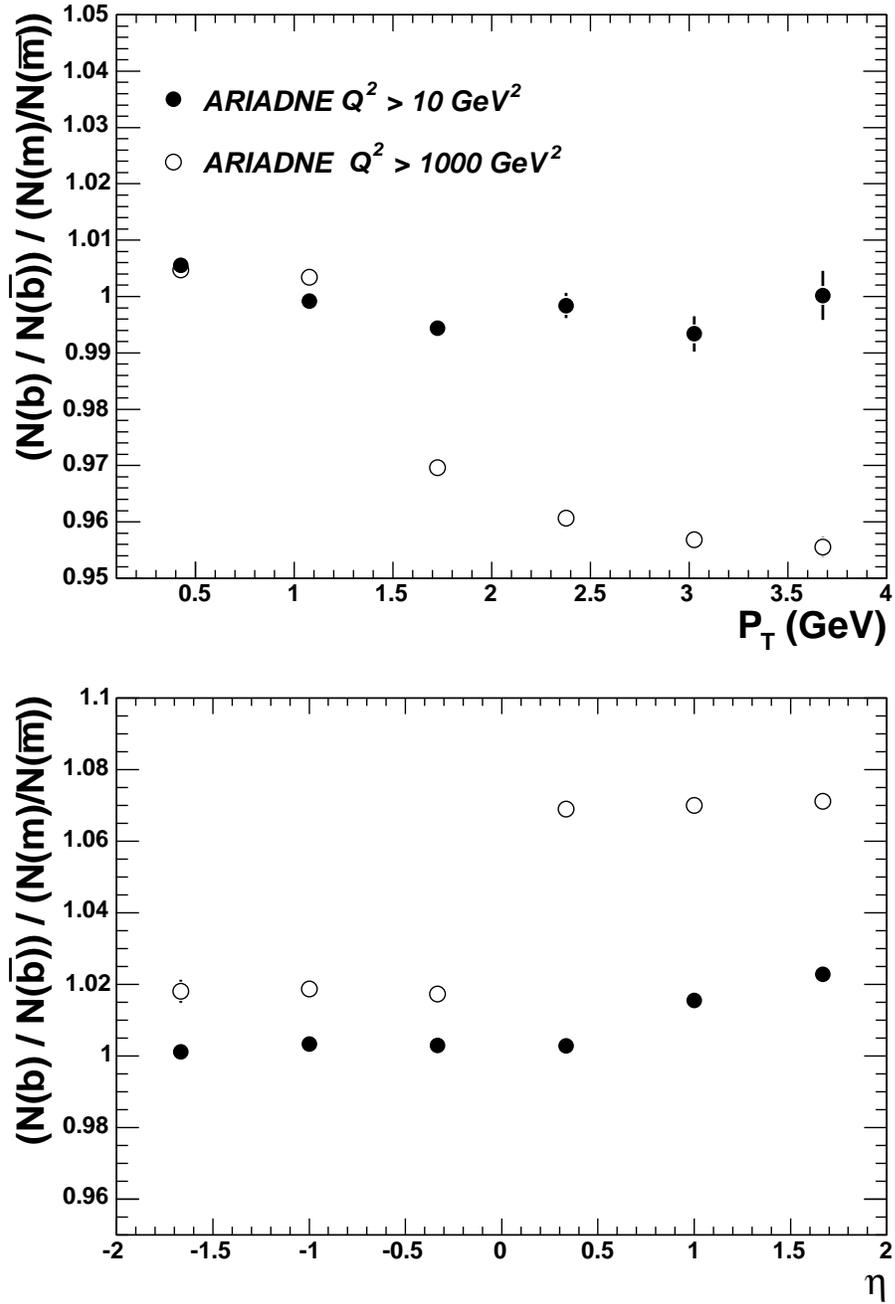,height=18cm}}
\caption
{
The double ratio (see the text)  
as functions of $p_T$ and $\eta$ for ARIADNE. 
The ratios are shown for two regions in
$Q^2$.
}
\label{rr_dis}
\end{center}
\end{figure}

To illustrate the presence of the net baryon number from the incoming beam better, 
one can calculate the double ratio, $r(b)/r(h)$.  
This double ratio has the advantage that the trivial effects
related to leading hadrons in the current jet are removed, thus
$r(b)/r(h)$ is more sensitive to the presence of the baryonic enhancement
due to the proton-remnant fragmentation.
Fig.~\ref{rr_dis} illustrates such double ratios for ARIADNE, using the points shown in
Fig.~\ref{r_dis}. 
The result shows a significant deviation from unity.  
The effect is at the level of $2\%$ for $\eta\simeq 2$ at $Q^2>10\gev^2$. 
For $Q^2>1000\gev^2$, 
the double ratio can be as big as $7\%$  for the same $\eta$ region.
The observed effect is mainly due to baryons produced at low $p_T$.

The same studies were also performed with LEPTO and HERWIG (not shown). 
For Figs.~\ref{r_dis} and ~\ref{rr_dis}, the LEPTO model generated 
for $Q^2>10\gev^2$  with the soft colour interactions 
is somewhat above the ARIADNE model for  $Q^2>10\gev^2$, but it is below the ARIADNE
predictions for $Q^2>1000\gev^2$. HERWIG shows a stronger asymmetry than LEPTO 
for both the ratio and the double ratio, 
which is likely to be related to differences between the Lund and
cluster hadronisation models. 
fragmentation mechanism. Thus the ratios shown
in Figs.~\ref{r_dis} and ~\ref{rr_dis} 
have large model-dependent uncertainties. 

\begin{figure}
\begin{center}
\mbox{\epsfig{file=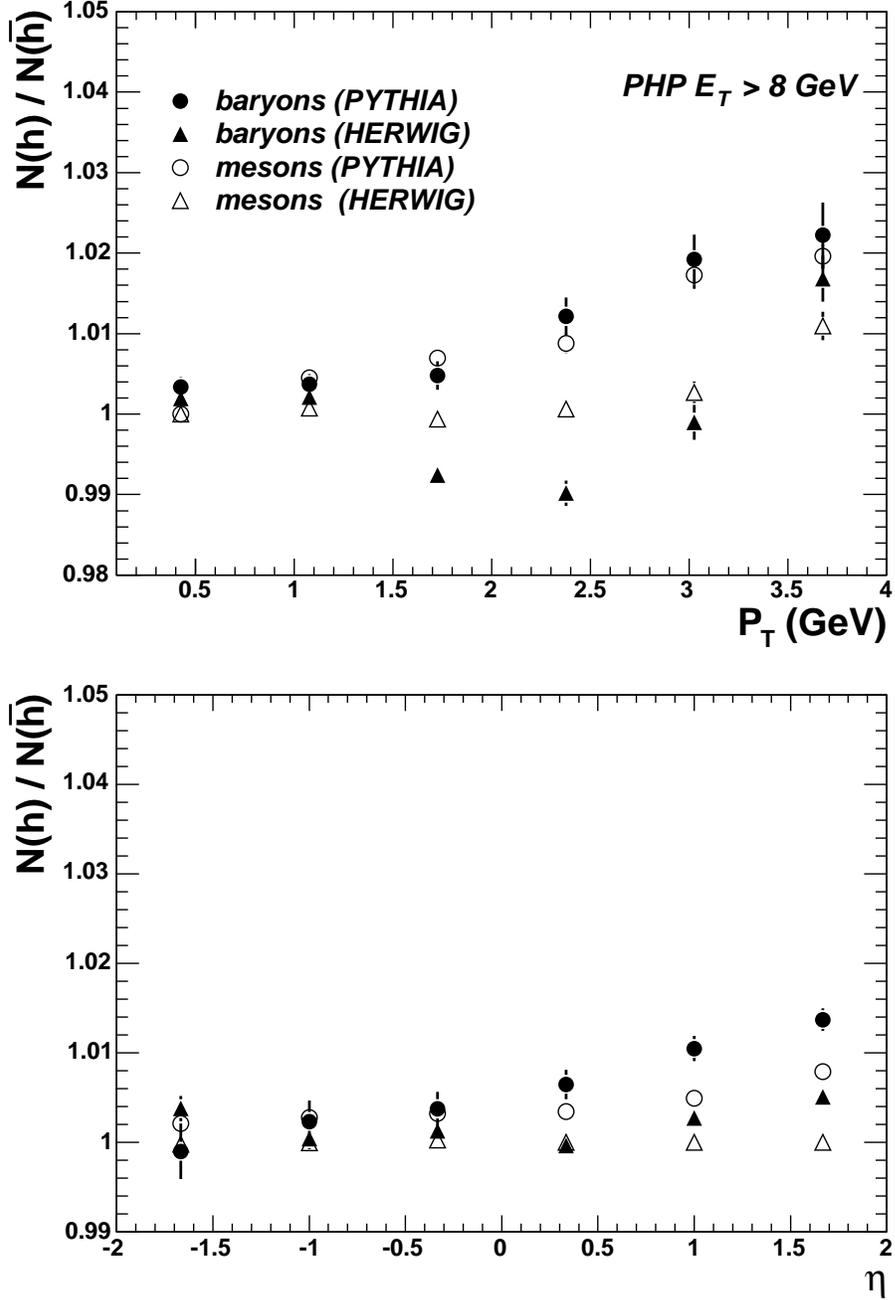,height=18cm}}
\caption
{
The ratio of the total number of final-state hadrons to antihadrons
as functions of $p_T$ and $\eta$. The  photoproduction events
were generated with PYTHIA 6.2 and  HERWIG 6.5 using the minimum
transverse momentum for the hard scale $E_T^{min}=8\gev$. 
The ratios are shown for mesons and baryons separately.
}
\label{r_php}
\end{center}
\end{figure}

\begin{figure}
\begin{center}
\mbox{\epsfig{file=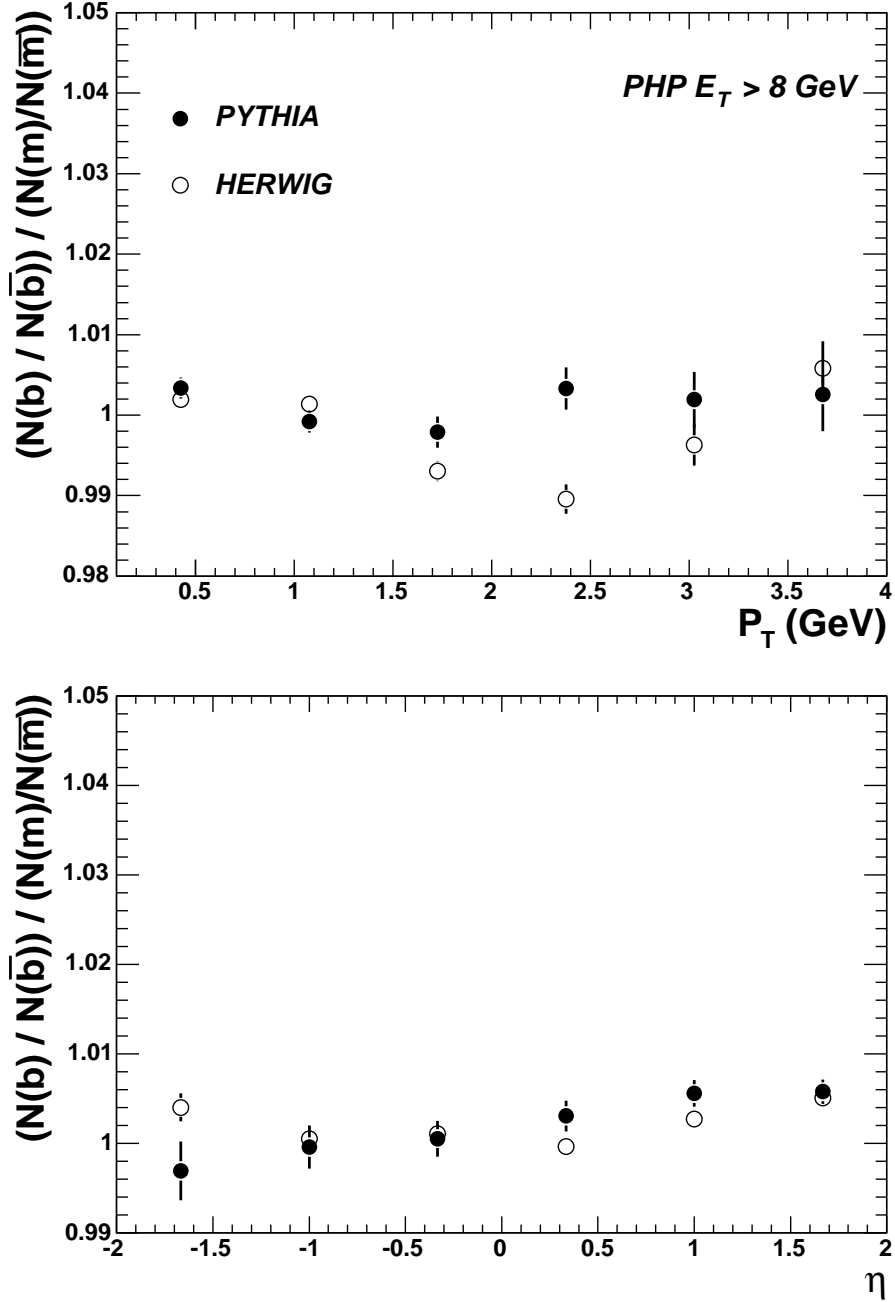,height=18cm}}
\caption
{
The double ratio (see the text)
as functions of $p_T$ and $\eta$ for photoproduction using the minimum
transverse momentum for the hard scale $E_T^{min}=8\gev$. 
Monte Carlo studies were performed using PYTHIA 6.2 and HERWIG 6.5.
}
\label{rr_php}
\end{center}
\end{figure}

Figures~\ref{r_php} and \ref{rr_php} show similar ratios for 
photoproduction events obtained using
PYTHIA and HERWIG. Both resolved and direct contributions
were included in the simulation. The events were
generated by requiring the minimum transverse 
momenta $E_T^{min}=8\gev$ for the hard subprocess.
The  asymmetry shown in Fig~\ref{r_php} is significantly smaller than for DIS.
For the double ratio, the effect is below $0.5\%$ for $-2<\eta<2$.
   
Thus, the MC simulations indicate that, for DIS events, a contribution from
the fragmentation of the proton remnant is present
even in the central fragmentation region, $\mid \eta \mid <1.5$, 
i.e. where the $\tp$ state was observed by ZEUS. 
This observation has a potential to 
explain the presence of the $\tp$ state as well as other 
possible exotic baryonic states 
in ZEUS DIS data~\cite{Chekanov:2004kn,Chekanov:2004ku}. 
At the same time, this mechanism leads to  
a reduction of the $\tp$ production rate at low $Q^2$ and  photoproduction,
which is consistent with the ZEUS studies of $\tp$ 
peak\footnote{Another 
possible explanation of the reduction of $\tp$ signal
at low $Q^2$ was discussed in~\cite{Chekanov:2004kn}.}. 
Since there is no strong evidence for antipentaquarks, the proposed
explanation is not in contradiction with the ZEUS data. 

The baryon-antibaryon asymmetry can also be studied by reconstructing 
$\Lambda / \bar{\Lambda}$, $p/\bar{p}$ etc. ratios.
For the $\Lambda / \bar{\Lambda}$ ratio \cite{Chekanov:2004ku}, 
there is indeed a large asymmetry,
but it is difficult to decouple it if from secondary-scattering interactions.
Recent study of $\Xi$ baryons \cite{Chekanov:2004ku} shows 
that the current statistical precision of HERA-I data  
is not yet sufficient to establish the $\Xi / \bar{\Xi}$ 
asymmetry at the $1-2\%$ level.
One should again note that the baryon-antibaryon ratio
measured in the central pseudorapidity region cannot give 
ultimate proof of the enhanced baryon production due to 
fragmentation of the proton remnant,
because a similar asymmetry should also be expected for all hadrons (including mesons)
due to leading hadrons in the current jet.  
As was discussed before, the  double ratio is more appropriate tool to
observe the discussed effect, since
the contribution from leading hadrons in the current jet is removed.

In summary,
if the pentaquark production is related to the 
proton-remnant fragmentation,
then pentaquarks should be 
best seen at low $p_T$ in the laboratory frame, and for positive values of $\eta$.
The production of antipentaquarks should be suppressed relative to pentaquarks.
In addition, high-$Q^2$ region is more favourable, as the incoming proton receives  
a stronger kick from the virtual photon; in the LUND 
fragmentation picture, a colour string 
stretched between the diquark and the struck quark has a higher chance for dragging
the diquark system to low values of pseudorapidity than for low $Q^2$ events. 
To verify this mechanism, high statistics data from HERA-II are necessary in order to
perform more differential studies of the $\tp$ signal.

%%%%%%%%%%%%%%%%%%%%%%%%%%%%%%%%%%%%%%%%%%%%%%%%%%%%%%%%%%%%%%%%%%
\section{Experimental reconstruction}
%%%%%%%%%%%%%%%%%%%%%%%%%%%%%%%%%%%%%%%%%%%%%%%%%%%%%%%%%%%%%%%%%%

It is noteworthy that, from the experimental point of view,
the decay channel $\tp \to \ksp$ 
is not simple since  
the $dE/dx$ method is often used  
to identify the protons, while $\ks$ are reconstructed
using the  secondary-vertex algorithms. In this approach, 
the identification of protons is  
only possible at low-momenta ($p<1.5\gev$ in ZEUS case).
However, in this case,  the phase space available for $\tp$ decay 
is dramatically reduced as the decay kinematics requires that 
the momenta  of the proton candidates 
should be typically higher than the momenta of $\ks$ \cite{Levchenko:2004sm}.
Note, in colliding experiments, a significant fraction
of $\ks$ mesons have momenta above $p\simeq 1.5\gev$, which is the
ZEUS cut used 
for the proton identification. 

The HERA data \cite{Airapetian:2003ri,Chekanov:2004kn} indicates
that the $\ksp$ invariant-mass spectrum has non-negligible
contributions from not well established PDG $\Sigma$ states.
Such baryons are not implemented in MC simulations, therefore,
the MC predictions (scaled to the luminosity of the data)
are significantly below the data for the $\ksp$ spectrum.
Such sensitivity
to the $\Sigma$ states may also be used by current and future experiments
to test the sensitivity to $\tp$.
 
The resolution for the
$\ksp$ invariant mass is another central problem for pentaquark
searches as the width of the $\tp$ baryon is predicted to be narrow.
ZEUS is among the experiments which have excellent resolution for this study,
$\simeq 2.5\mev$ \cite{Chekanov:2004kn}
(using the PDG mass for $\ks$ to improve the 
resolution for the $\ksp$ mass distribution).
Keeping in mind the difficult
background near the $1480\mev$ mass region due to the PDG $\Sigma(1480)$ bump,
a structure near the $1520\mev$ region will be difficult to observe if the
resolution would be $3-4$ times larger.

As was discussed before, if the contribution of the 
proton remnant to the central
fragmentation region is expected at high $Q^2$, then one
may expect that the pentaquark production is also enhanced at high $Q^2$.   
However, from the experimental point of view, 
pentaquark studies at very high $Q^2$ could be affected by 
high combinatorial background.
This will be discussed in more details below.

%%%%%%%%%%%%%%%%%%%%%%%%%%%%%%%%%%%%
\subsection{Combinatorial background and experimental sensitivity}
%%%%%%%%%%%%%%%%%%%%%%%%%%%%%%%%%%%%

At high energies, the combinatorial background
for the invariant-mass reconstruction 
scales approximately as $<n>^2$, where $n$ is the number of 
produced tracks. 
Going to a high energy ($\sqrt{s}$, jet $E_T$, $Q^2$, $W$, etc.) available
for hadroproduction,   
combinatorics for fully-inclusive
events is rising due to a large production rate 
of non-heavy flavour hadrons
produced by soft fragmentation and resonance decays.  
For HERA, an increase in the 
track multiplicity and combinatorial background is expected
for high-$E_T$ photoproduction events, or for 
very high $Q^2$ events in DIS.     
This should mask the peaks 
from hadrons which may carry the  
quantum numbers of colliding beams,  
or states which can only be produced after the fragmentation 
of partons created by hard subprocesses.

The most relevant example illustrating this point 
is heavy-flavour mesons or baryons.
The production of such states by pure fragmentation
is significantly suppressed, but they can be produced in  
the hard BGF subprocess $\gamma^{*}g\to c\bar{c}$ at the fragmentation stage.
Thus, for high-energy processes 
(or for high-$E_T$ jets), 
there is a stronger combinatorial background than at low energy, 
since a significant fraction of strange kaons and  baryons   
produced by 
hadronisation and resonance decays can contribute to the relevant invariant-masses distributions. 

This point has been checked by using the same MC samples as those shown 
in Figs.~\ref{r_dis}-\ref{rr_php}.
First, the decay channel  $D^0\to K^-\pi^+$ was reconstructed
by combining the momenta of two final-state charged particles.
It was found that, for the ARIADNE event sample generated 
at $Q^2>10\gev^2$, the ratio
of the signal events over the background 
was about $2.3$, 
while for the photoproduction sample of PYTHIA with $E_T^{min}=8\gev$, 
this ratio was  $1.5$.  
Such difference is mainly related to a high particle multiplicity in photoproduction
data than in DIS. 
In contrast, when $\phi(1020)$ mesons decaying to $K^+K^-$ are reconstructed, 
the signal-over-background ratios
were $1.3$ and $1.2$ for DIS and photoproduction, respectively.     
Thus, the signal-over-background ratio 
for the states which can be produced by pure fragmentation (like $\phi$-mesons) 
is about the same for DIS and photoproduction, while it is significantly
different for hadrons which have a production mechanism driven by the hard
subprocess or by other mechanisms unrelated to pure fragmentation and
resonance decays.

In this respect, the statement \cite{Dzierba:2004db} that  
experiments with superior statistics
for known hadrons (such as $\Lambda(1520)$) 
have better chances to find $\tp$ is not fully
correct for high-energy experiments.
The production mechanism of such states is soft fragmentation,
thus the signal-over-background ratio does not change significantly
as a function of the energy available for the hadronic final state. 
In contrast, the production mechanism
of the $\tp$ state is unlikely related to pure fragmentation mechanism,  
thus the combinatorial background for the $\tp$ 
reconstruction is expected to increase with energy.
Therefore, high  statistics for reconstructed 
non-heavy flavour mesons and baryons 
can only indicate the quality of particle identification. This is
an important issue, but not the most important 
compared to possibly exotic production mechanism of the
$\tp$ state, and the experimental difficulties related to 
the increase of combinatorial background with energy.

Note that for the  
$\Lambda(1520)$ baryons, which have a similar decay mode $K^-p$ as the
$\tp$ state, 
in some cases it is also difficult to say 
about the experimental sensitivity to the $\tp$, 
even if one assumes the same production
mechanism as for the pentaquark state. 
Indeed, the ZEUS preliminary results \cite{Chekanov:2004hd,*Karshon:2004kt}
showing the $\Lambda(1520)$
signal at $Q^2>20\gev^2$ does not look impressive compared to 
the corresponding peak from BaBar~\cite{Aubert:2004bm}.
This is due to the fact that the reconstruction of $\Lambda(1520)$ at ZEUS 
was done in a significantly smaller allowed
phase space than that for the $\tp$, since $K^{\pm}$ mesons 
were reconstructed by using the $dE/dx$, i.e. 
only $K^{\pm}$ mesons with relatively low momenta,  
$p<0.8\gev$, were used. In contrast, the $\ks$ sample 
used for the $\tp$ searches did not have such limitation,  
since no the $dE/dx$ requirement was used (there is only
$20\%$ of $\ks$ in the range $p<0.8\gev$). The protons
used for the $\Lambda(1520)$ reconstruction  
were also selected by using a significantly stronger $dE/dx$ cut.

It seems that the best known state which can be used to
check the experimental 
sensitivity to $\tp$, and to compare it with other experiments,  
is $\Lambda_c$ (using the $\ksp$ decay channel).   
For $ep$ collisions, 
the  $\Lambda_c$ baryons are produced through the BGF mechanism, 
$\gamma^{*}g\to c\bar{c}$.  
Thus the cross section for $\Lambda_c$ is decreasing with increase of
Bjorken $x$ and $Q^2$. Unfortunately,
if the $\tp$ production mechanism proposed in Sect.~\ref{prod} is
correct, the yield of $\Lambda_c$ baryons decreases for 
the kinematic regions where the $\tp$ cross section
is expected to be high.

%%%%%%%%%%%%%%%%%%%%%%%%%%%%%%%%%%
\subsection{Particle misidentification}
%%%%%%%%%%%%%%%%%%%%%%%%%%%%%%%%%%%

Particle misidentifications in hadron spectroscopy 
can roughly be divided into the following categories: 
1) misinterpretation of a peak which may correspond to a known
state when a different mass assumption for the decay products is used, and 2)
misreconstruction of a state 
after using the same tracks twice (track splitting). 
We  will not discuss here the first problem 
as  this is always a first check to be made to exclude possible 
reflections  (which usually
never produce a very sharp peak consistent with the detector resolution).   

Here we will concentrate on the second problem.
It was already pointed 
out by several authors~\cite{Zavertyaev:2005yf} 
(see other details in \cite{Dzierba:2004db}) 
that, for example, using ghost tracks associated to 
$\Lambda$-baryon decays can lead to peaks at $1540\mev$.

Generally, the experiments which have reported the $\tp$ state are
well aware of the danger related to the ghost tracks.
The issue is probably more important for high-energy experiments
with large combinatorial background under the observed peak as it is impossible to 
scan each event by eye in order to verify which track belongs to which decaying state.

Note that all experiments which observe 
the $\ksp$ peak claim the reconstructed $\tp$ masses somewhere 
between $1520$ and $1530\mev$,
so the spurious peak shown in \cite{Dzierba:2004db} is by $20-10\mev$ shifted
from the observed peak.   
At ZEUS, everything was done to avoid such 
double counting by removing 
$\Lambda$ hyperons from the consideration and by 
reducing the background under the $\ks$
peak as much as possible.
The contribution from the $\Lambda$ decays was further reduced
using the tracks fitted to the primary vertex for the proton candidates.
     
Finally, ZEUS observes the $\tp$ peak only for 
a specific type of events, mainly for DIS.
It is hard to imagine why the spurious peak appears in DIS only, 
but not for high-$E_T$ photoproduction events which have
a high average track multiplicity and a larger    
production rate of $\Lambda$ baryons 
compared to DIS.
 
The last observation is probably a good hint for 
future pentaquark searches: 
if a signal disappears for certain class of events 
(where it can  potentially disappear
due to some physical reasons, and 
assuming that such class of events was not preselected using
track-related quantities), 
this would likely mean that the observed peak is not
due to track-related misreconstructions.

%%%%%%%%%%%%%%%%%%%%%%%%
\subsection{Conclusions}
%%%%%%%%%%%%%%%%%%%%%%%

The HERA collider is unique and plays a 
significant role in pentaquark searches. 
This uniqueness is in a possible access to the 
net baryon number from the colliding proton 
beam, even when the most central
fragmentation region is used for the baryon reconstruction.
This conclusion is mainly relevant for DIS, while the
effect is significantly smaller for gluon-driven photoproduction.
If the assumptions made in this paper are correct, the pentaquark
signal should be best seen in the forward region of pseudorapidity,
at low $p_T$ and at medium or high $Q^2$ regions. The signal from
antipentaquarks should be suppressed compared to pentaquarks.    

Another feature of $ep$ collisions at HERA is relatively low energy for 
the hadronic final state compared 
to the Tevatron experiments. This has a clear advantage as it
leads to a low combinatorial 
background for the $\ksp$ mass reconstruction.  
Furthermore, according to several 
predictions~\cite{Titov:2004wt,*Diakonov:2004ie}, 
the $\tp$ production
cross section can be suppressed at high energies.   

For the nearest future, HERA  could 
be the only high-energy colliding experiment which
is well suited for the exotic baryon searches, since 
the pentaquark production in $\ee$ annihilation is expected
to be suppressed compared to the experiments with baryons
in the colliding beams. Therefore, the future Linear Collider
is likely to be not well suited for such searches.    
From the other hand, the hadronic final state in $pp$ collisions
at LHC is expected to be 
too complicated, and a significant effort should be made to 
deal with the combinatorial background for the $\ksp$ invariant mass. 

\section*{Acknowledgements}
\vspace{0.3cm}

I thank many of my colleagues from the ZEUS, H1, HERMES and HERA-B experiments.   
Especially I would like to 
thank R.~Yoshida and U.~Karshon for the discussions and comments on earlier versions 
of the manuscript.

%%%%%%%%%%%%%%%%%%%%%% references %%%%%%%%%%%%%%%%%%%%%%%%%%%%%%
% draft type, with title and hp-ex
% \bibliographystyle{./Macros/h-elsevier-clean}
\bibliographystyle{./Macros/l4z_pl}
\def\bibname{\Large\bf References}
\def\refname{\Large\bf References}
\pagestyle{plain}
\bibliography{biblio}

\providecommand{\etal}{et al.\xspace}
\providecommand{\coll}{Collaboration}
\catcode`\@=11
\def\@bibitem#1{%
\ifmc@bstsupport
  \mc@iftail{#1}%
    {;\newline\ignorespaces}%
    {\ifmc@first\else.\fi\orig@bibitem{#1}}
  \mc@firstfalse
\else
  \mc@iftail{#1}%
    {\ignorespaces}%
    {\orig@bibitem{#1}}%
\fi}%
\catcode`\@=12
\begin{mcbibliography}{10}

\bibitem{Nakano:2003qx}
LEPS~Collaboration, T.~Nakano, et~al.,
\newblock Phys. Rev. Lett.{} 91~(2003)~012002\relax
\relax
\bibitem{Barth:2003ja}
SAPHIR~Collaboration, J.~Barth, et~al.,
\newblock Phys. Lett.{} B572~(2003)~127\relax
\relax
\bibitem{Stepanyan:2003qr}
CLAS~Collaboration, S.~Stepanyan, et~al.,
\newblock Phys. Rev. Lett.{} 91~(2003)~252001\relax
\relax
\bibitem{prl92:032001}
CLAS~Collaboration, V.~Kubarovsky, et~al.,
\newblock Phys. Rev. Lett.{} 92~(2004)~032001\relax
\relax
\bibitem{Barmin:2003vv}
DIANA~Collaboration, V.~V. Barmin, et~al.,
\newblock Phys. Atom. Nucl.{} 66~(2003)~1715\relax
\relax
\bibitem{Aleev:2004sa}
SVD~Collaboration, A.~Aleev, et~al.,
\newblock Preprint \mbox{hep-ex/0401024}, 2004\relax
\relax
\bibitem{hep-ex-0403011}
COSY-TOF~Collaboration, M.~Abdel-Bary, et~al.,
\newblock Phys. Lett.{} B~595~(2004)~127\relax
\relax
\bibitem{Airapetian:2003ri}
HERMES~Collaboration, A.~Airapetian, et~al.,
\newblock Phys. Lett.{} B~585~(2004)~213\relax
\relax
\bibitem{Chekanov:2004kn}
ZEUS~Collaboration, S.~Chekanov, et~al.,
\newblock Phys. Lett.{} B~591~(2004)~7\relax
\relax
\bibitem{zp:a359:305}
D.~Diakonov, V.~Petrov, M.~V. Polyakov,
\newblock Z.~Phys.{} A~359~(1997)~305\relax
\relax
\bibitem{Bai:2004gk}
BES~Collaboration, J.~Z. Bai, et~al.,
\newblock Phys. Rev.{} D~70~(2004)~012004\relax
\relax
\bibitem{Abt:2004tz}
HERA-B~Collaboration, I.~Abt, et~al.,
\newblock Phys. Rev. Lett.{} 93~(2004)~212003\relax
\relax
\bibitem{Antipov:2004jz}
SPHINX~Collaboration, Y.~M. Antipov, et~al.,
\newblock Eur. Phys. J.{} A21~(2004)~455\relax
\relax
\bibitem{pl:599:1}
ALEPH~Collaboration, S.~Schael, et~al.,
\newblock Phys. Lett.{} B~599~(2004)~1\relax
\relax
\bibitem{cdfpenta}
CDF~Collaboration, D.~O. Litvintsev,
\newblock {\em Presented at 6th International Conference on Hyperons, Charm and
  Beauty Hadrons (BEACH 2004)}.
\newblock Chicago, Illinois, USA (2004).
\newblock Also in preprint \mbox{FERMILAB-CONF-04-205-E}
  (\mbox{hep-ex/0410024})\relax
\relax
\bibitem{Aubert:2004bm}
BABAR~Collaboration, B.~Aubert, et~al.,
\newblock {\em Submitted to 32nd International Conference on High-Energy
  Physics (ICHEP 04)}.
\newblock Beijing, China (2004).
\newblock Also in preprint \mbox{hep-ex/0408064}\relax
\relax
\bibitem{Abe:2004ws}
BELLE~Collaboration, R.~Mizuk,
\newblock {\em Proc. of International Workshop on PENTAQUARK04}.
\newblock Spring-8, Hyogo, Japan (2004).
\newblock Also in preprint \mbox{hep-ex/0411005}\relax
\relax
\bibitem{Chekanov:2004qm}
ZEUS~Collaboration, S.~Chekanov, et~al.,
\newblock Eur. Phys. J.{} C~38~(2004)~29\relax
\relax
\bibitem{Chekanov:2004ku}
ZEUS~Collaboration, S.~Chekanov, et~al.,
\newblock Preprint \mbox{hep-ex/0501069}, 2005\relax
\relax
\bibitem{Chekanov:2004hd}
S.~Chekanov,
\newblock {\em Proc.\ DIS04 Workshop}, D.~Bruncko, J.~Ferencei,
  P.~Str\'{i}\v{z}enec~(eds.), Vol.~2, p.~576.
\newblock IEP SAS, Strbsk\'e Pleso,~Slovakia (2004)\relax
\relax
\bibitem{Karshon:2004kt}
U.~Karshon,
\newblock {\em Presented at International Workshop on PENTAQUARK04 (Spring-8)}.
\newblock Hyogo, Japan (2004).
\newblock Also in preprint \mbox{hep-ex/0410029}\relax
\relax
\bibitem{Aktas:2004qf}
H1~Collaboration, A.~Aktas, et~al.,
\newblock Phys. Lett.{} B~588~(2004)~17\relax
\relax
\bibitem{Dzierba:2004db}
A.~R. Dzierba, C.~A. Meyer, A.~P. Szczepaniak,
\newblock Preprint \mbox{hep-ex/0412077}, 2004\relax
\relax
\bibitem{Hicks:2004ge}
K.~Hicks,
\newblock Preprint \mbox{hep-ex/0412048}, 2004\relax
\relax
\bibitem{Kabana:2005kg}
S.~Kabana,
\newblock Preprint \mbox{hep-ph/0501121}, 2005\relax
\relax
\bibitem{Adloff:1998yg}
H1~Collaboration, C.~Adloff, et~al.,
\newblock Eur. Phys. J.{} C6~(1999)~587\relax
\relax
\bibitem{Adloff:2001be}
H1~Collaboration, C.~Adloff, et~al.,
\newblock Nucl. Phys.{} B619~(2001)~3\relax
\relax
\bibitem{Chekanov:2002pf}
ZEUS~Collaboration, S.~Chekanov, et~al.,
\newblock Nucl. Phys.{} B~637~(2002)~3\relax
\relax
\bibitem{Chekanov:2002yh}
ZEUS~Collaboration, S.~Chekanov, et~al.,
\newblock Nucl. Phys.{} B~658~(2003)~3\relax
\relax
\bibitem{Sjostrand:2001yu}
T.~Sj\"{o}strand, L.~L\"onnblad, S.~Mrenna,
\newblock Preprint \mbox{hep-ph/0108264}, 2001\relax
\relax
\bibitem{herwig}
G.~Corcella, et~al.,
\newblock JHEP{} 0101~(2001)~10\relax
\relax
\bibitem{Lonnblad:1992tz}
L.~L\"onnblad,
\newblock Comput. Phys. Commun.{} 71~(1992)~15\relax
\relax
\bibitem{Ingelman:1996mq}
G.~Ingelman, A.~Edin, J.~Rathsman,
\newblock Comput. Phys. Commun.{} 101~(1997)~108\relax
\relax
\bibitem{runmc}
S.~Chekanov,
\newblock Preprint \mbox{hep-ph/0411080}, 2004,
\newblock available on \texttt{http://www.desy.de/\til chekanov/runmc}\relax
\relax
\bibitem{hep-ph/0003255}
S.~Chekanov,
\newblock {\em Proc. of the workshop ``Monte Carlo Generators for HERA
  Physics''}, A.~Doyle, et~al.~(eds.), p.~309.
\newblock DESY, Hamburg, Germany (1999).
\newblock Also in preprint \mbox{hep-ph/0003255}\relax
\relax
\bibitem{Levchenko:2004sm}
B.~B. Levchenko,
\newblock Preprint \mbox{hep-ph/0401122}, 2004\relax
\relax
\bibitem{Zavertyaev:2005yf}
M.~Zavertyaev,
\newblock Preprint \mbox{hep-ex/0501028}, 2005\relax
\relax
\bibitem{Titov:2004wt}
A.~Titov, A.~Hosaka, S.~Date, Y.~Ohashi,
\newblock Phys. Rev.{} C~70~(2004)~042202\relax
\relax
\bibitem{Diakonov:2004ie}
D.~Diakonov,
\newblock Preprint \mbox{hep-ph/0406043}, 2004\relax
\relax
\end{mcbibliography}

%%%%%%%%%%%%%%%%%%%%%%%%%%%%%% FIGURES %%%%%%%%%%%%%%%%%%%%%%%%%%%%%%

%%%%%%%%%% PHP

\end{document}